\documentclass[twocolumn,superscriptaddress,showpacs,preprintnumbers,amsmath,amssymb,oneside,prl]{revtex4}
\usepackage{graphicx}
\usepackage{dcolumn}
\usepackage{bm}
\usepackage{amssymb}
\usepackage{amsmath}
\usepackage{amsfonts}
\usepackage{amssymb}
\usepackage[bookmarks = false]{hyperref}
\usepackage{color}

\begin{document}

\title{Freezing motion-induced dephasing in an atomic-ensemble quantum memory}
\author{Yan Jiang}
\affiliation{Hefei National Laboratory for Physical Sciences at Microscale and Department
of Modern Physics, University of Science and Technology of China, Hefei,
Anhui 230026, China}
\affiliation{CAS Center for Excellence and Synergetic Innovation Center in Quantum Information and Quantum Physics, University of Science and Technology of China, Hefei, Anhui 230026, China}
\author{Jun Rui}
\affiliation{Hefei National Laboratory for Physical Sciences at Microscale and Department
of Modern Physics, University of Science and Technology of China, Hefei,
Anhui 230026, China}
\affiliation{CAS Center for Excellence and Synergetic Innovation Center in Quantum Information and Quantum Physics, University of Science and Technology of China, Hefei, Anhui 230026, China}
\author{Xiao-Hui Bao}
\affiliation{Hefei National Laboratory for Physical Sciences at Microscale and Department
of Modern Physics, University of Science and Technology of China, Hefei,
Anhui 230026, China}
\affiliation{CAS Center for Excellence and Synergetic Innovation Center in Quantum Information and Quantum Physics, University of Science and Technology of China, Hefei, Anhui 230026, China}
\author{Jian-Wei Pan}
\affiliation{Hefei National Laboratory for Physical Sciences at Microscale and Department
of Modern Physics, University of Science and Technology of China, Hefei,
Anhui 230026, China}
\affiliation{CAS Center for Excellence and Synergetic Innovation Center in Quantum Information and Quantum Physics, University of Science and Technology of China, Hefei, Anhui 230026, China}
\date{\today}
\begin{abstract}
Motion-induced dephasing is a dominant decoherence mechanism for atom-gas quantum memories. In this paper, we develop a new coherent manipulation technique which enables arbitrary engineering of the spin-wave momentum with neglectable noise. By zeroing the spin-wave momentum, motion-induced dephasing can be frozen completely. We experimentally demonstrate this scheme with laser-cooled atoms in a DLCZ configuration. By applying the freezing pulses, memory lifetime gets extended significantly to the limit of atom cloud expansion and does not depend on the detection angle anymore. The observed high cross-correlation above 20 proves that high-fidelity memory operation is well preserved after coherent manipulation.
\end{abstract}

\maketitle


Atomic ensembles are a promising approach for quantum memory~\cite{Sangouard2011,Hammerer2010,Bussieres2013}. Compared with single-particle quantum memories~\cite{Blinov2004,Specht2011,Gao2012,Maurer2012}, atomic ensembles are advantageous due to collective enhancement which enables efficient interaction with single-photons~\cite{Duan2001}. Successful implementation of atomic-ensemble quantum memory has been carried out in many different physical systems~\cite{Sangouard2011,Hammerer2010,Bussieres2013}, such as laser cooled atoms, atoms in a vapor cell, and rare-earth ions doped in a solid crystal.
With atom-gas ensembles, remarkable progresses have been made in recent years, such as the realization of a quantum repeater node~\cite{Chou2007,Yuan2008}, demonstration of quantum teleportation between memories~\cite{Bao2012pnas,Krauter2013}.

In an atomic-ensemble quantum memory, a single-photon is stored as a collective excitation (spin-wave)~\cite{Duan2001,Fleischhauer2005}, which can be described as
\begin{equation*}
|g...g...g\rangle\Rightarrow |\Psi\rangle_{gs}=\frac{1}{\sqrt{N}}\sum_{j}^{N}e^{i\mathbf{k}_{s}\cdot
\mathbf{r}_{j}}|g...s_{j}...g\rangle ,
\end{equation*}
where $|g\rangle$ is the initial state for all atoms, $|s\rangle$ is another state for storage, $N$ denotes number of atoms, and $\mathbf{r}_{j}$ denotes the position of the $j$-th atom in the ensemble. During storage, the momentum mismatch between the input photon and the control field is also imprinted on to the spin-wave~\cite{Bo2009}, which is denoted as $\mathbf{k}_{s}$ in the above collective state. The spin-wave momentum determines the single-photon emission direction during retrieval. Atom motions during storage will randomize the internal phases and ruin the collective interference process during retrieval, which becomes a dominant decoherence mechanism for atom-gas quantum memories~\cite{Bo2009}. The motion-induced dephasing can be suppressed significantly by employing a collinear configuration~\cite{Bo2009,Wang2014}, in which case $\mathbf{k}_{s}\approx0$. Nevertheless, leakage from the collinear control beam into the single-photon channel is extremely difficult to suppress, and requires using multiple stages of filters, which will reduce the overall memory efficiency significantly. Motion-induced dephasing can also be eliminated by using optical lattice to confine atom motion in the spin-wave direction~\cite{Ran2009,Radnaev2010,Dudin2013}. However, optical lattice is technically demanding and gets the experimental setup complicated. Moreover, by making coherent manipulations between $|g\rangle$ and $|s\rangle$, motion-induced dephasing can also get rephased~\cite{Rui2015}. In the experimental demonstration~\cite{Rui2015}, noise due to manipulation errors was suppressed to below the single-photon level, nevertheless high-fidelity storage and retrieval require much better Raman manipulation techniques.

In this paper, we introduce a new coherent manipulation technique by making use of an auxiliary state of $|s'\rangle$, which enables significant suppression of the noise due to manipulation errors. Moreover, by configuring the Raman beam directions, we can arbitrarily engineer the spin-wave momentum. By setting $\mathbf{k}_{s'} =0$, motion-induced dephasing can be frozen completed. In our experimental demonstration, we make use of a laser-cooled atomic ensemble memory and perform the DLCZ protocol. High nonclassical correlation between the write-out and read-out photon is preserved. We also demonstrate the lifetime extension process for a series of detection angles. Without applying the freezing pulses, memory lifetime depends on the detection angle significantly. After applying the freezing pulses, memory lifetime extends to a constant value which is limited by thermal expansion.

\begin{figure}[hbtp]
\includegraphics[width=\columnwidth]{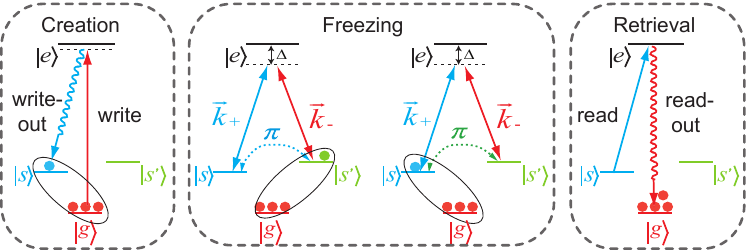}
\caption{(color online). Experimental scheme of spin-wave freezing. A single spin-wave $|\Psi\rangle_{gs}$ is first created in the atomic ensemble between the states $|g\rangle\leftrightarrow|s\rangle$. Afterwards, a Raman $\pi$ pulse which couples the $|s\rangle\leftrightarrow|s'\rangle$ transition via a two-photon Raman process is then applied immediately to transfer the spin-wave to $|\Psi\rangle_{gs'}$  with zero wave vector. Later, just before spin-wave retrieval, another $\pi$ pulse is employed to restore the spin-wave back to $|\Psi\rangle_{gs}$. Finally the spin-wave is retrieved back to a read-out photon by turning on the read beam. The black ellipse refers to the spin-wave states.}
\label{Fig1}
\end{figure}

The new spin-wave manipulation scheme is depicted in Fig.~\ref{Fig1}. The two ground states $|g\rangle$ and $|s\rangle$, together with an excited state $|e\rangle$, form a $\Lambda$-type system. In the DLCZ protocol~\cite{Duan2001}, a write pulse creates a write-out photon and a single atomic spin-wave simultaneously through spontaneous Raman scattering. After some duration of storage, the spin-wave is later retrieved as a read-out photon. High cross-correlation between the write-out and the read-out photon guarantees high-fidelity memory operations. The spin-wave wave-vector is determined by the write-out photon detection angle $\theta_s$ relative to the write beam, as $\mathbf{k}_{s}=\mathbf{k}_{w}-\mathbf{k}_{wo}$. For a typical non-collinear configuration, motion-induced dephasing gives rise to a lifetime of tens to hundreds of $\mu$s~\cite{Bo2009}. To manipulate the spin-wave wave-vector without introducing strong noise~\cite{Rui2015}, we make use of an auxiliary ground state $|s'\rangle$, and two Raman beams with wave-vectors $\mathbf{k}_{+}$ and $\mathbf{k}_{-}$ to couple the transition between $|s\rangle$ and $|s'\rangle$ through a two-photon stimulated Raman transition \cite{Mark1991}. Immediately after the spin-wave being created, a Raman $\pi$ pulse is applied to convert the spin-wave $|\Psi\rangle_{gs}$ to another one $|\Psi\rangle_{gs'}$ with a wave-vector of $\mathbf{k}_{s'}=\mathbf{k}_{s}+\mathbf{k}_{\pi}$, where $\mathbf{k}_{\pi}=\mathbf{k}_{+}-\mathbf{k}_{-}$. By carefully arranging the Raman beams such that $\mathbf{k}_{s'}=0$, the momentum of the spin-wave is erased, and then the atomic-motion-induced dephasing is frozen out. After a certain duration of storage, another $\pi$ pulse is turned on to restore the spin-wave $|\Psi\rangle_{gs'}$ back to $|\Psi\rangle_{gs}$ and gives an additional wave vector $\mathbf{k'}_{\pi}=\mathbf{k}_{-}-\mathbf{k}_{+}$. Thus the original spin-wave wave-vector $\mathbf{k}_{s}$ is recovered, which enables subsequently efficient retrieval along the original phase-matching direction. Since population in the $|g\rangle$ state is unaffected, imperfection of the $\pi$ pulses between $|s\rangle$ and $|s'\rangle$ will only result in slight loss for the stored spin-wave, which will not increase the readout noise level, so high non-classical correlation can be well preserved.

\begin{figure}[hbtp]
\includegraphics[width=\columnwidth]{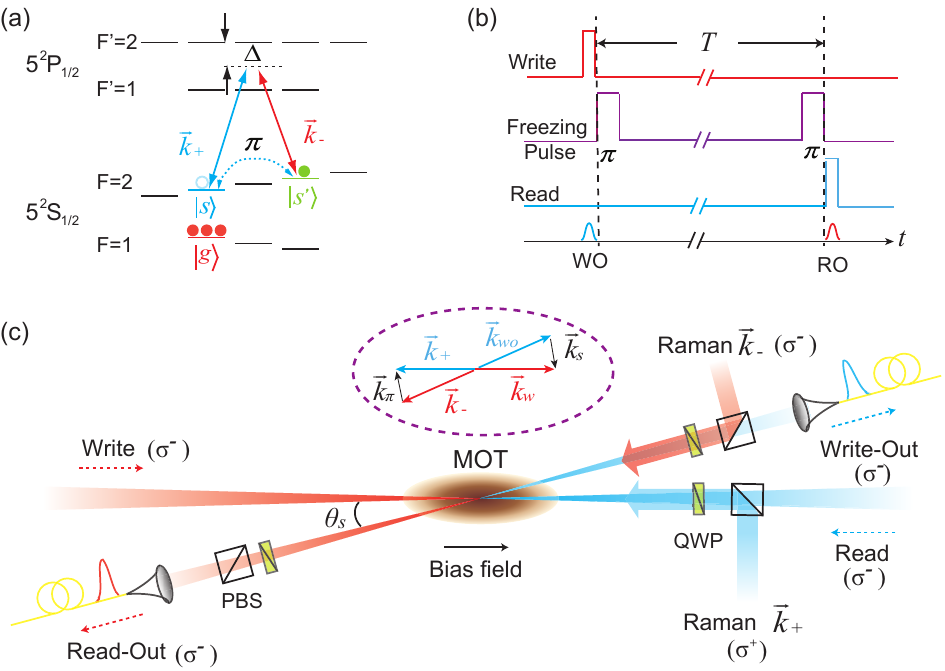}
\caption{(color online). Experimental process and setup. (a) Energy levels and Raman beam configuration. $|g\rangle$ and $|s'\rangle$ are selected as a pair of ``clock states". (b) Time sequence of the experiment. The write and read pulses have a duration of 100 ns. The Raman $\pi$ pulses have a duration of about 2 $\mu$s. The first Raman $\pi$ pulse is turned on immediately after the write process. $T$ is the storage duration. (c) Experimental layout. The write beam couples the transition $|F=1\rangle\leftrightarrow|F'=2\rangle$ with a detuning of -10 MHz and a power of 1 $\mu$W. The read pulse resonantly couples the $|F=2\rangle\leftrightarrow|F'=2\rangle$ transition with a power of 100 $\mu$W. The waist of the write/read beam is 200 $\mu$m. For the write-out/read-out detection mode, the waist is 90 $\mu$m. A bias magnetic field is applied to define the quantization axis and split the Zeeman levels. In order to freeze out the motion-induced decoherence, intersection angle between the two Raman beams is arranged to be exactly the same as $\theta_s$. The write-out and read-out photons are first polarization filtered before single-mode fiber coupling, and then frequency filtered with etalons and detected with single photon counters (not shown). QWP stands for a quarter-wave plate and PBS stands for a polarized beam-splitter. Inset: Momentum relationships for write, write-out and Raman beams.}
\label{Fig2}
\end{figure}

Our experimental setup is shown in Fig.~\ref{Fig2}. An atomic ensemble of $\sim10^{8}$ atoms is loaded by a magneto-optical trap (MOT) and result in a temperature of about 13 $\mu$K after polarization gradient cooling. The optical depth is about 5. A bias magnetic field of 1.5 G is applied to lift the Zeeman degeneracy. The energy levels are carefully selected as $|g\rangle=|F=1, m_{F}=-1\rangle$, $|s\rangle=|F=2, m_{F}=-1\rangle$ and $|s'\rangle=|F=2, m_{F}=+1\rangle$ of the hyperfine ground states of $^{87}$Rb, so that $|g\rangle$ and $|s'\rangle$ are a pair of ``clock states'', which is insensitive to inhomogeneity of magnetic field during storage. Consequently, with this configuration, two dominant decoherence mechanisms, atomic motions as well as inhomogeneity of magnetic field, can be suppressed simultaneously.

To drive the transition between $|s\rangle$ and $|s'\rangle$, two Raman beams which originate from the same diode laser are applied. To employ constructive interference between the transition pathes via $|F'=2\rangle$ and $|F'=1\rangle$, single-photon detuning of the Raman beams is set to be $\Delta=-408$ MHz relative to $|F'=2\rangle$, which is exactly half of the hyperfine splitting between the $D1$ excited states.
Each Raman beam has a power of about $2.0$ mW. To achieve high-fidelity Raman pulses, the diameter of the Raman beams is set to 3.8 mm to guarantee the intensity flatness in the atom region, and the intensities of both Raman beams are actively stabilized during the MOT loading phase and then passively hold during the experiment phase. Ramsey measurements are carried out to optimize the two-photon detuning to zero. The measured Raman Rabi frequency is about $230$ kHz, leaving the width of each $\pi$ pulse to be about 2.0 $\mu$s. To fully freeze the motion-induced dephasing, the wave vectors should strictly satisfy $(\mathbf{k}_{w}-\mathbf{k}_{wo})+(\mathbf{k}_{+}-\mathbf{k}_{-})=\mathbf{0}$ , which means that the Raman $\mathbf{k}_{+}$ should either propagate along the write-out beam, or counter-propagate along the write beam. But restricted by the optical helicities and effective filters, here we can only arrange the Raman $\mathbf{k}_{+}$ and $\mathbf{k}_{-}$ beams as shown in Fig.~\ref{Fig2}(c), and the momentum relationships between write, write-out, and Raman beams are shown in the insert. Two stage AOMs are used to reduce the leakage of Raman $\mathbf{k}_{-}$ beam into the read-out channel. What's more, the frequency of  Raman $\mathbf{k}_{-}$ beam is about $6.8$ GHz different from the read-out photons, and thus the noise from this beam to the read-out channel can be further reduced significantly by the etalon filtering.

\begin{figure}[hbtp]
\includegraphics[width=\columnwidth]{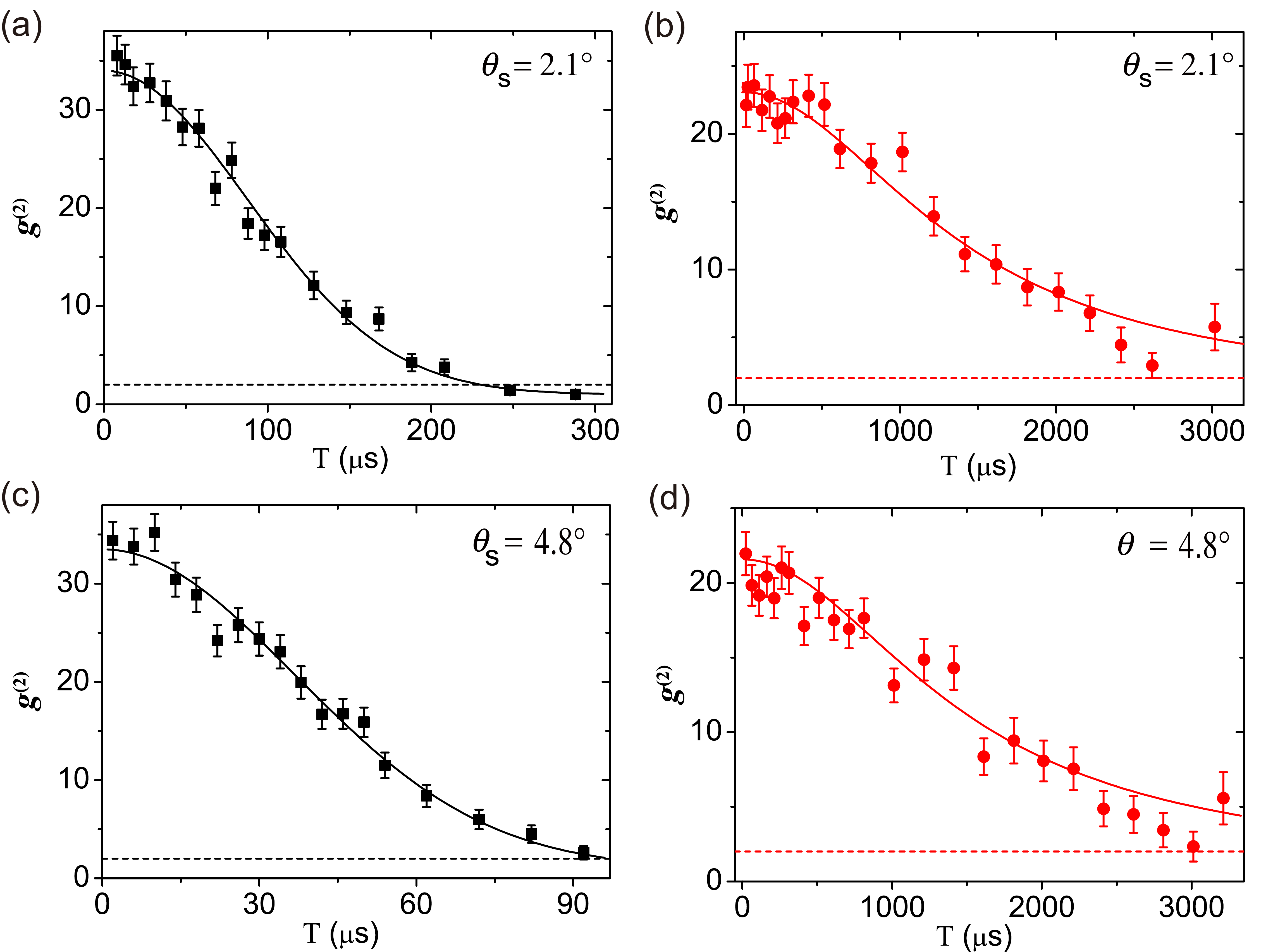}
\caption{(color online).  Measured cross-correlation as a function of storage time. (a) and (c), Raman manipulation is not applied. (b) and (d), Raman manipulation is applied. The lifetime fitting models used are the same as \cite{Bo2009}.}
\label{Fig3}
\end{figure}

Below we carry out DLCZ storage to investigate the feasibility of our spin-wave freezing scheme. Detailed time sequences and geometric layout for the control beams and detection modes are shown in Fig.~\ref{Fig2}.
The quality of DLCZ storage is typically characterized by the cross-correlation function between the write-out and read-out photons, which is defined as $g^{(2)}=p_{\rm{wo,ro}}/(p_{\rm{wo}}p_{\rm{ro}})$, where $p_{\rm{wo}}(p_{\rm{ro}})$ denotes the probability of detecting a write-out (read-out) photon, and $p_{\rm{wo,ro}}$ is the coincident probability between a write-out and a read-out photon. $g^{(2)}>2$ means the read-out photon is non-classically correlated with the write-out photon~\cite{Kuzmich2003}. We first measure decay of $g^{(2)}$ without applying the Raman pulses, with the results shown in Fig.~\ref{Fig3}(a) and (c) for two different storage angles of $\theta_s=2.1^{\circ}$ and $\theta_s=4.8^{\circ}$. In these measurements, the write-out probability is set to be $p_{\rm{wo}}\simeq0.38\%$. The measured initial $g^{(2)}$ are well above 30, which implies high degree of nonclassical correlations between the write-out and the read-out photons. When applying the Raman freezing pulses, we get new experimental results as shown in Fig.~\ref{Fig3}(b) and (d) under the same level of $p_{\rm{wo}}$. The measured initial $g^{(2)}$ are 23.05$\pm$0.53 and 20.6$\pm$0.56 respectively, which implies high non-classical correlations have been well preserved. If such a DLCZ interface is employed to generated a pair of atom-photon entanglement\cite{Chen2007b}, the visibility can be well approximated as $V\simeq(g^{(2)}-1)/(g^{(2)}+1)$~\cite{Riedmatten2006}, and thus the entanglement fidelity can be estimated as $F\simeq(1+3V)/4\simeq(g^{(2)}-0.5)/(g^{(2)}+1)$~\cite{White2007}. As the measured $g^{(2)}$ after Raman manipulations are well above 20, a corresponding high fidelity of $F\simeq0.93$ can be estimated for entanglement generation. In comparison, our previous spin echo experiment~\cite{Rui2015} got a measured $g^{(2)}\simeq5$ after Raman manipulation, which corresponds to a fidelity of $F\simeq0.75$ for entanglement generation. Notably, for both storage angles, $g^{(2)}$ drops slightly after the freezing process, which may be caused by the imperfect $\pi$ pulses and the imperfect mode matching during Raman manipulations. With a finite $\pi$ pulse fidelity, there will be some losses for stored spin-wave state after two $\pi$ pulses, which will reduce the retrieval efficiency and get $g^{(2)}$ dropped. Since we cannot realize $\mathbf{k}_{\pi}=-\mathbf{k}_s$ perfectly, there will still be a certain degree of motional dephasing during storage, which will also cause drops of $g^{(2)}$.

\begin{figure}[hbtp]
\includegraphics[width=\columnwidth]{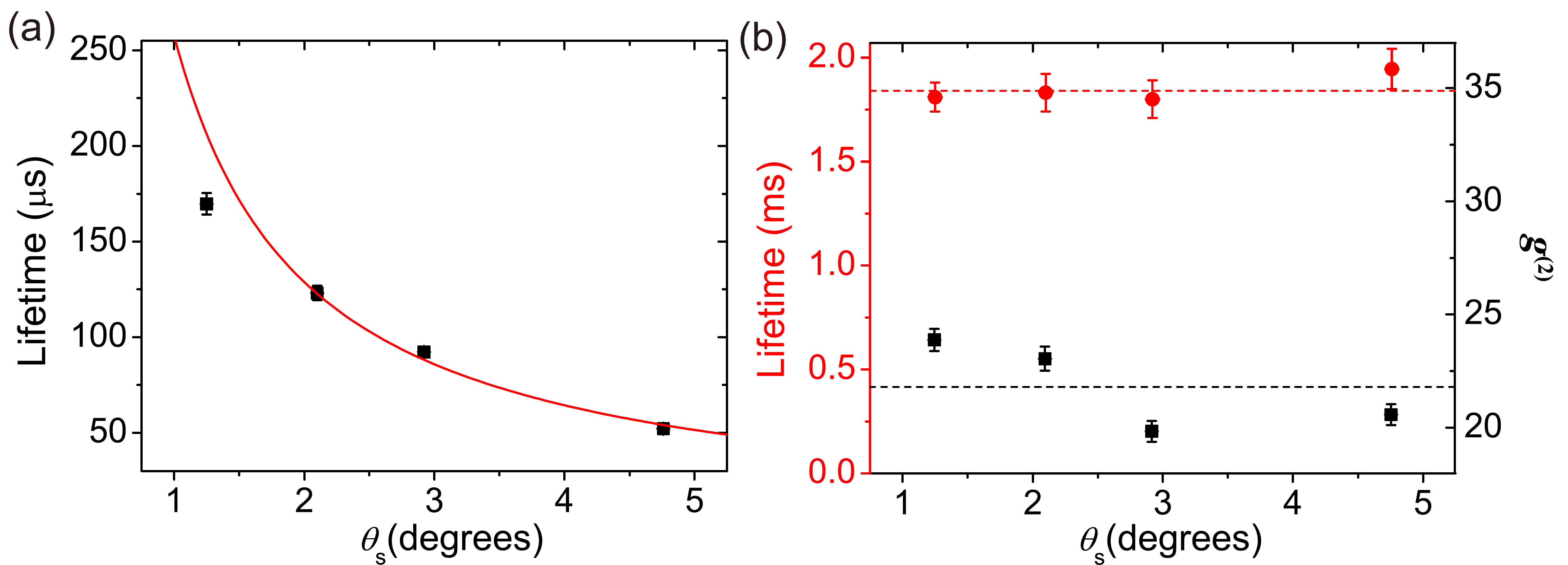}
\caption{(color online). Comparison for DLCZ storages with and without freezing for different storage angles. (a), Without freezing, the measured lifetime is inversely proportional to $\theta_{s}$ as expected. The red curve is the theoretical estimation under the formula $\tau_{s}=1/k_{s}v$, with $v=\sqrt{\pi k_{B}T/2m}$ being the one-dimensional speed of the atoms and $T$ is the temperature of the ensemble. (b), Lifetime (red circle) and initial cross-correlation (black square) with freezing. The measured lifetime does not depend on the storage angle anymore due to the elimination of the motion-induced dephasing and mean lifetime is about $1.84$ ms, which is shown as a dashed red line. Cross-correlation is well preserved with a mean value of 21.8 shown as a dashed black line.}
\label{Fig4}
\end{figure}

Moreover, by applying the Raman freezing pulses, memory lifetime gets extended significantly as also shown comparatively in Fig.~\ref{Fig3}. Without Raman manipulation, the fitted $1/e$ lifetime is 123(1) $\mu$s for $\theta_s=2.1^{\circ}$ and 52(1) $\mu$s for $\theta_s=4.8^{\circ}$. After applying the freezing pulses, $1/e$ lifetime is increased to 1.83(1) ms and 1.94(5) ms, respectively. Furthermore, we perform the same lifetime measurement for more storage angles, and summarize all results in Fig.~\ref{Fig4}. Without spin-wave freezing (shown in Fig.~\ref{Fig4}(a)), the storage lifetime is inversely proportional to the detection angle $\theta_{s}$ as theoretically estimated \cite{Bo2009}. 1/e lifetime varies from 169 $\mu$s to 52 $\mu$s as the detection angle $\theta _{s}$ increases from $\theta _{s}=1.25^{\circ}$ to $\theta _{s}=4.8^{\circ}$. To be mentioned that, when storage angle close to or smaller than $1^{\circ}$, despite of the motion-induced dephasing, inhomogeneity of magnetic field will be another main decoherence mechanism to restrict the storage lifetime, so the measured lifetime for $\theta_s=1.25^{\circ}$ is slightly smaller than the theoretical one. As a comparison, for storages with freezing, the lifetime is no longer sensitive to the storage angles anymore, as shown in Fig.~\ref{Fig4}(b), leaving a constant lifetime of about 1.84 ms, which is close to the thermal expansion limit \cite{Bo2009}. What's more, the measured $g^{(2)}$ values are all around 20, which guarantees high-fidelity memory operations.

In summary, we have introduced and experimentally demonstrated a new coherent manipulation technique for atomic ensemble quantum memories. By transferring the atomic excitation to an auxiliary state, we can arbitrarily engineer the spin-wave momentum. By minimizing the spin-wave momentum, we freeze the motion-induced dephasing completely and extend the memory lifetime significantly to the limit of atom cloud expansion. Since population inversion is not involved, high non-classical correlations above 20 have been observed between the write-out and read-out photons in the DLCZ scheme, which guarantees that high-fidelity memory performance is well preserved after spin-wave manipulation. Our work enriches the experimental toolbox of harnessing atomic ensembles for high-performance quantum memories, especially for holographic quantum memories where many spin-waves with different wave-vectors are used \cite{Dai2012}.

This work was supported by the National Natural Science Foundation of China, the Chinese Academy of Sciences, and the National Fundamental Research Program of China. X.-H.B. specially acknowledge support from the Youth Qianren Program.

\bibliography{freezing}

\end{document}